\documentclass[sigconf]{acmart}
\renewcommand\footnotetextcopyrightpermission[1]{}
\usepackage{booktabs}
\usepackage{tabularx}
\usepackage{array}
\usepackage{microtype}
\usepackage{hyperref}
\usepackage{enumitem}
\usepackage{graphicx}
\usepackage{tikz}
\usepackage{xcolor}
\usetikzlibrary{positioning, shapes.geometric, arrows.meta, fit, backgrounds}
\usepackage{fancyhdr}

\setcopyright{none}
\acmDOI{}
\acmISBN{}
\acmConference[]{}{}{}
\acmYear{2026}
\acmPrice{}

\begin{document}

\title[GreenZ: A Sustainable UX Framework]{GreenZ: A Sustainable UX Framework\\for Complex Digital Systems}

\author{Trisha Solanki}
\affiliation{%
  \institution{UXperiment Inc.}
  \city{British Columbia}
  \country{Canada}
}
\email{support@uxperiment.design}
\renewcommand{\shortauthors}{Solanki T.}

\begin{abstract}
Digital systems have become simultaneously more powerful and more
wasteful. Features accumulate that nobody uses. Data is collected that
nobody analyzes. AI is deployed at significant energy and water cost for
gains that a simpler approach could have achieved. And through all of
it, the people who depend on these systems quietly absorb the
consequences in cognitive load, lost time, and eroded trust.

Existing frameworks address pieces of this. Blevis~\cite{blevis2007}
established sustainable interaction design as a formal practice.
The W3C Web Sustainability Guidelines~\cite{w3cwsg2023} address
front-end efficiency. Amershi et al.~\cite{amershi2019} and Google
PAIR~\cite{googlepair2021} offer responsible AI interaction guidance.
But no published framework integrates environmental sufficiency,
cognitive sufficiency, and AI necessity into a single operational
system for enterprise product teams.

This paper introduces \textbf{GreenZ}, a three-layer Sustainable UX
Framework for complex digital systems. Its three layers are a
Philosophy Layer built around ten published principles, an Operational
Frameworks Layer comprising five applied systems, and a Tools and
Canvases Layer of practical audit instruments and decision models. Two
contributions sit at the framework's core: a \textbf{Digital Waste
Taxonomy} classifying eight distinct waste types, and an \textbf{AI
Sufficiency Decision Model} that asks \textit{whether} AI should exist
in a given flow before any question of \textit{how} to implement it.

GreenZ v1 is theoretically grounded but empirically unvalidated. A
practitioner expert review study is underway at time of submission. The
paper presents the framework's architecture, its conceptual foundations,
its position relative to existing literature, and an honest account of
what remains to be established.
\end{abstract}


\maketitle

\section{Introduction}

The digital economy generates waste at a scale that is only beginning
to be measured honestly. McGovern's synthesis of industry data
reports that roughly 90\% of collected data is never accessed after
three months~\cite{mcgovern2020}. Pendo's 2019 Feature Adoption
Report, drawing on usage data from 615 enterprise software
subscriptions, found that 80\% of product features were rarely or
never used~\cite{pendo2019}. Luccioni, Jernite and Strubell measured
88 AI models and found that the least efficient image generation model
consumed energy equivalent to charging 522 smartphones, for a single
image~\cite{luccioni2024}. Li et al.\ estimate that training a
language model at the scale of GPT-3 can evaporate 700,000 litres of
clean freshwater for data centre cooling~\cite{li2024}.

These numbers have a common cause. Systems are designed without
sufficiency as a first-class constraint. Features are added because
they are possible. Data is captured because infrastructure is cheap.
AI is deployed because it is available. The cost falls on users in
cognitive load and lost time, on infrastructure in energy and water,
and on organizations in eroded trust and mounting maintenance burden.

Sustainable HCI has documented this pattern for nearly two decades.
Blevis~\cite{blevis2007} established Sustainable Interaction Design
by linking invention to material disposal. Mankoff et
al.~\cite{mankoff2007} drew the foundational distinction between
sustainability \textit{through} design, changing user behaviour, and
sustainability \textit{in} design, reducing the artifact's footprint.
Knowles, Bates and H{\aa}kansson~\cite{knowles2018} made the field's
sharpest argument: that ecological sustainability requires contraction,
while the prevailing economic model requires expansion, and that
sustainable HCI must engage with that structural tension rather than
optimize around it.

Despite this body of work, Silberman et al.\ admitted in 2014 that
sustainable HCI had demonstrated ``little impact outside
HCI''~\cite{silberman2014}. The problem is structural. Academic
sustainable HCI operates at a level of theory that does not translate
directly into product decisions. Practitioner frameworks - Nielsen's
heuristics, Google's HEART, Microsoft's HAX Toolkit, Google PAIR's
Guidebook - are, without exception, sustainability-blind. They
optimize interaction quality without asking whether the feature, the
data collection, or the AI inference was needed at all.

GreenZ is designed to work in the space between these two
traditions. It is neither a purely academic framework nor a
practitioner checklist. It is a three-layer knowledge system intended
to be rigorous enough to cite and operational enough to use on a
Monday morning.

\section{Background and Related Work}

\subsection{Sustainable HCI}

The field was formally named at CHI 2007 by Mankoff, Blevis, Borning,
Friedman, Fussell, Hasbrouck, Woodruff and Sengers~\cite{mankoff2007},
who proposed the in/through-design distinction that still structures the
literature. Blevis's concurrent SID paper introduced rubrics for
material effects - disposal, salvage, recycling, remanufacturing, reuse,
longevity, sharing, heirloom status, active repair - and five design
principles connecting invention to end-of-life.

DiSalvo, Sengers and Brynjarsdo\'{o}ttir~\cite{disalvo2010} identified
five genres of SHCI research and a persistent problem: most work treats
sustainability as an application domain, something to design
\textit{for}, rather than as a paradigm shift that should reshape how
design is done. That tension between instrumentalist and transformative
approaches has never been fully resolved, and GreenZ takes a position
on it: the framework is instrumentalist in its tools and transformative
in its philosophy, deliberately so.

Knowles, Bates and H{\aa}kansson~\cite{knowles2018} published the
field's sharpest reckoning, arguing that sustainability cannot be
achieved through efficiency gains within existing growth systems.
Santarius et al.~\cite{santarius2023} later formalized ``digital
sufficiency'' across four dimensions - hardware, software, user, and
economic - which provides GreenZ's primary theoretical grounding for
the Design Degrowth principle.

\subsection{Digital Waste}

``Digital waste'' is not yet a formalized academic construct. The
closest published treatments are scattered across several literatures.
McGovern~\cite{mcgovern2020} aggregated industry statistics on unused
data, feature abandonment, and app attrition, making the case that
digital is a hidden accelerant of climate harm. Gray, Kou, Battles,
Hoggatt and Toombs~\cite{gray2018}, Mathur et al.~\cite{mathur2019},
and Gray, Santos and Bielova~\cite{gray2024} developed dark pattern
taxonomies, now codified in EU Digital Services Act Article 25 and
FTC guidance~\cite{ftc2022}, that describe intentional UX waste:
friction designed to extract value rather than deliver it.
Hwang~\cite{hwang2020} framed attention economics as a speculative
bubble of inflated engagement metrics.

GreenZ synthesizes these threads into a unified Digital Waste Taxonomy
(Section~\ref{sec:taxonomy}), treating waste not as a single phenomenon
but as eight distinct types with different causes, costs, and
remediation paths. To the authors' knowledge, no published framework
has previously consolidated this literature into a single named taxonomy
for product-team use.

\subsection{Complexity in Enterprise Systems}

Norman's distinction between complexity and
complication~\cite{norman2010}, complexity is inherent structure,
complication is confusion produced by bad design, frames the enterprise
UX problem well. His ``Simplicity Is Not the Answer''~\cite{norman2008}
is important here: enterprise systems genuinely are complex, and
pretending otherwise produces interfaces that hide necessary structure
rather than surface it.

Endsley's three-level Situation Awareness model~\cite{endsley1995}, 
Perception, Comprehension, Projection, provides the cognitive
architecture that enterprise UX implicitly needs but rarely names.
Systems that fail at any of the three levels generate errors not because
users are incompetent but because the interface does not support what
the user is actually trying to do.

Klein's Recognition-Primed Decision model~\cite{klein1999}, derived
from studies of firefighters, ICU nurses, and military commanders,
established that expert decision-making under time pressure is
pattern-based, not comparative. Experts recognize situations and
mentally simulate their first plausible action. Enterprise UX built
around option menus and comparison patterns actively disrupts this
process by forcing deliberative cognition onto a recognition task.

\subsection{Responsible AI UX}

Amershi et al.'s 18 Guidelines for Human-AI Interaction~\cite{amershi2019}
represent the most-cited practitioner framework in this space. The
guidelines address \textit{how} to design AI interactions across four
phases. They presuppose that the AI feature has already been decided.

Google PAIR's People~+~AI Guidebook~\cite{googlepair2021} includes a
``Determine if AI adds value'' section, the closest existing pattern to
an AI necessity question, but it functions as a heuristic prompt, not
a decision model. Microsoft's HAX Toolkit~\cite{microsoft2021}
operationalizes Amershi et al.'s guidelines but similarly assumes AI
presence as given.

Crawford's \textit{Atlas of AI}~\cite{crawford2021} provides the
critical counterpoint: AI systems are material infrastructure with
planetary extraction costs. Crawford and Joler's ``Anatomy of an AI
System''~\cite{crawfordjoler2018} mapped the full lifecycle cost of a
conversational AI device. Luccioni et al.'s FAccT 2024 study provided
the quantitative basis for treating AI inference as a resource event
rather than a free computation~\cite{luccioni2024}.

The gap GreenZ addresses is distinct from all of the above: no
published framework asks \textit{whether} AI should exist in a given
flow as its primary question. GreenZ's AI Sufficiency Decision Model
(Section~\ref{sec:ai}) does.

\subsection{The Accessibility--Sustainability Bridge}

W3C's Web Accessibility Initiative frames accessibility, usability, and
inclusion as overlapping concerns. The W3C Web Sustainability
Guidelines~\cite{w3cwsg2023}, modelled on WCAG, extend this to
environmental sustainability. Watson~\cite{watson2024} notes that W3C
is now positioning sustainability as a fifth horizontal principle
alongside accessibility, security, privacy, and internationalization.

The mechanical overlap between accessibility and sustainability is
significant and underexplored. Semantic HTML reduces JavaScript payload.
Plain language reduces cognitive load. Progressive loading reduces both
data transfer and time-to-comprehension. GreenZ treats these overlaps
as opportunities, not coincidences.

\section{Framework Architecture}

GreenZ is organized in three layers, each serving a distinct function.

\textbf{Layer~1 Philosophy} answers: \textit{Why does this matter and
what do we believe?}

\textbf{Layer~2 Operational Frameworks} answers: \textit{How do we
act on those beliefs in practice?}

\textbf{Layer~3 Tools and Canvases} answers: \textit{What does a
practitioner pick up and use today?}

Each layer is independently useful. A team can adopt a single audit
tool without committing to the philosophical layer. An organization
building internal standards can adopt the operational frameworks without
using UXperiment's canvases. This modularity is intentional: it
supports adoption at whatever pace a given organization can sustain,
without requiring a big-bang commitment that most product teams cannot
make.

The layers are not equally visible to the frameworks GreenZ is designed
to sit alongside. Nielsen's heuristics, HEART, and Amershi et al.\ all
operate at Layer~2 or~3. GreenZ adds a Layer~1 that asks \textit{why}
before any question of \textit{how}.

\begin{figure}[h]
\centering
\begin{tikzpicture}[
  every node/.style={font=\small},
  layer/.style={
    rectangle, rounded corners=4pt,
    text width=\columnwidth-24pt,
    minimum height=1.1cm,
    align=center,
    draw=black!60,
    line width=0.6pt
  },
  arrow/.style={
    -Stealth,
    line width=1pt,
    color=black!50
  }
]

\node[layer, fill=black!8] (l1) {
  \textbf{Layer 1: Philosophy}\\[2pt]
  \footnotesize GreenZ Principles \quad Digital Waste Theory\\
  Human-Centred Complexity \quad Responsible Intelligence
};

\node[layer, fill=black!4, below=0.35cm of l1] (l2) {
  \textbf{Layer 2: Operational Frameworks}\\[2pt]
  \footnotesize SUX Audit \quad AI Sufficiency \quad Complexity Reduction\\
  Metrics Framework \quad Responsible Workflow
};

\node[layer, fill=white, below=0.35cm of l2] (l3) {
  \textbf{Layer 3: Tools \& Canvases}\\[2pt]
  \footnotesize Audit Worksheets \quad Decision Trees \quad Sprint Templates\\
  Scoring Systems \quad Evaluation Checklists
};

\draw[arrow] (l1.south) -- (l2.north);
\draw[arrow] (l2.south) -- (l3.north);

\end{tikzpicture}
\caption{GreenZ three-layer architecture. Each layer is independently
usable; together they form a complete knowledge system from worldview
to working artifact.}
\label{fig:architecture}
\end{figure}
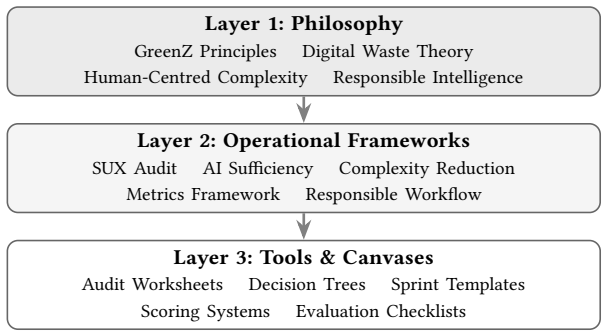

\section{Layer 1: Philosophy}

\subsection{The GreenZ Principles}
\label{sec:principles}

GreenZ's philosophical core is ten principles, published openly at
\url{uxperiment.design/greenz}. They are the worldview that produced
the operational frameworks, not a prerequisite for using them. Each
principle is grounded in existing theory; none is invented from scratch.

\vspace{4pt}
\noindent\textbf{1. Regenerative Interfaces.}
\textit{Every touchpoint gives back.}
Interfaces that restore cognitive clarity and reduce user fatigue
generate positive outcomes beyond the immediate transaction. This
principle draws on Kaplan and Kaplan's Attention Restoration
Theory~\cite{kaplan1989}: recovery from directed-attention fatigue
is a design deliverable, not a wellness afterthought.

\vspace{4pt}
\noindent\textbf{2. Waste-Conscious UX.}
\textit{No user journey is wasted. No click is trash.}
Dead-ends, unused features, and redundant interactions are treated as
digital waste with measurable costs in time, compute, and trust. This
principle operationalizes McGovern's~\cite{mcgovern2020} digital waste
argument and Pendo's~\cite{pendo2019} feature-adoption findings at the
level of individual design decisions.

\vspace{4pt}
\noindent\textbf{3. Adaptive Systems Thinking.}
\textit{Design for how systems behave, not just how they look.}
Complex systems are more resilient when they distribute function,
recover from failure gracefully, and evolve with use. This principle
draws on Alexander et al.'s pattern languages~\cite{alexander1977},
Simon's architecture of complexity~\cite{simon1962}, and Meadows's
systems thinking framework~\cite{meadows2008}. Natural systems serve
here not as metaphor but as structural evidence that sufficiency and
resilience are compatible design properties.

\vspace{4pt}
\noindent\textbf{4. Carbon Shadow Audits.}
\textit{What you don't see still matters.}
Carbon shadow extends carbon footprint accounting to invisible
emissions: unnecessary animations, heavy back-end calls, redundant
API round-trips, compute-intensive inference paths. This principle is
grounded in Luccioni et al.~\cite{luccioni2024} and Li et
al.~\cite{li2024}, and aligns with W3C WSG guidance on avoiding
unnecessary requests.

\vspace{4pt}
\noindent\textbf{5. Psychologically Regenerative Design.}
\textit{Protect the mind, preserve the planet.}
Cognitive capacity depletes under sustained load. Design that produces
burnout or information overload depletes a shared human resource. Sweller's
Cognitive Load Theory~\cite{sweller1988} established that extraneous
load, load produced by poor design rather than by the task
itself, is always removable. Endsley's Situation Awareness
model~\cite{endsley1995} adds that systems failing to support
perception, comprehension, or projection of future states generate
errors regardless of user skill. Edmondson's psychological safety
research~\cite{edmondson1999} completes the picture: reducing fear of
failure supports clearer thinking. GreenZ treats extraneous cognitive
load as waste: simultaneously a human-factors intervention and a
sustainability obligation.

\vspace{4pt}
\noindent\textbf{6. Cultural Ecosystem Equity.}
\textit{Designs that grow across geographies and generations.}
Systems requiring high bandwidth, new hardware, or a culturally
specific reading of iconography are systems that exclude. Inclusive
design is not an add-on; it is a prerequisite for any design that
claims sustainability credentials. This principle aligns with Watson et
al.'s Inclusive Design Principles~\cite{watson2019} and the CRPD's
framing of accessibility as a human right.

\vspace{4pt}
\noindent\textbf{7. Design Degrowth.}
\textit{Growth $\neq$ more screens. Progress = smarter choices.}
The enterprise software industry's default orientation - more features,
larger dashboards, expanded AI capabilities - produces systems that are
powerful in specification and exhausting in use. Design degrowth
proposes sufficiency as the primary design criterion. The degrowth
framing originates in political economy~\cite{latouche2009,hickel2020}
and was brought into sustainable HCI by Knowles, Bates and
H{\aa}kansson~\cite{knowles2018}. GreenZ applies it as a
design-process discipline, not a political position: an enterprise can
pursue growth while still asking, at every feature decision, whether
what it is about to build needs to exist.

\vspace{4pt}
\noindent\textbf{8. Time-to-Value Conservation.}
\textit{Sustainable design respects user time like it is a resource.}
User time is finite, non-renewable, and unequally distributed. Simon's
foundational observation that ``a wealth of information creates a
poverty of attention''~\cite{simon1971} established the basic economics.
Williams~\cite{williams2018} and Hwang~\cite{hwang2020} developed the
critique of systems designed to extract attention rather than respect
it. GreenZ connects this directly to resource economics: a user who
completes a task in three steps instead of seven generates fewer API
calls, fewer rendering cycles, and less infrastructure load. Attention
conservation and carbon conservation are, at the task level, the same
design goal. This principle is distinct from ``time well spent''
framing (Harris, 2014), which addresses individual wellbeing; this
principle addresses systemic and environmental cost.

\vspace{4pt}
\noindent\textbf{9. Regenerative Design Systems.}
\textit{Every pattern reused is one less built from scratch.}
Where Principle~1 concerns what users experience, this principle
concerns how the work is built and maintained. Design systems built
for longevity - modular, reusable, component-based - reduce redundant
effort across every product that inherits them. The environmental case
is direct: design work that does not need to be redone consumes no
additional compute, no review cycles, no handoff overhead. This
principle is grounded in Frost's atomic design~\cite{frost2016} and
Blevis's longevity and reuse principles~\cite{blevis2007}. It is
explicitly distinguished from circular economy theory~\cite{emf2013}:
the concern here is design artifact longevity within digital systems,
not physical material recovery.

\vspace{4pt}
\noindent\textbf{10. Speculative Sustainability R\&D.}
\textit{We invest in what is not invented yet.}
Responsible innovation requires designating organizational time for
exploratory work without immediate commercial application. This
principle is about organizational practice - the deliberate reservation
of capacity for non-billable sustainability research - not speculative
design as critical methodology~\cite{dunneraby2013}. Where Dunne and
Raby use fiction to interrogate the present, this principle uses
structured R\&D investment to build toward a different future.
Frameworks that do not evolve will be superseded; this one is designed
to evolve.

\subsection{Digital Waste Taxonomy}
\label{sec:taxonomy}

The framework introduces an eight-type taxonomy of digital waste,
synthesizing McGovern~\cite{mcgovern2020}, Gray et
al.~\cite{gray2018,gray2024}, Mathur et al.~\cite{mathur2019},
Pendo~\cite{pendo2019}, Santarius et al.~\cite{santarius2023},
Luccioni et al.~\cite{luccioni2024}, and Hwang~\cite{hwang2020}.

\begin{table}[h]
\caption{Digital Waste Taxonomy}
\label{tab:taxonomy}
\small
\begin{tabularx}{\columnwidth}{>{\bfseries}p{1.6cm} X p{1.6cm}}
\toprule
Type & Definition & Primary Cost \\
\midrule
Compute & AI inference, API calls, or rendering cycles that produce no
marginal user value & Energy, infra cost \\
\addlinespace
Data & Fields collected but never used in decisions or experience &
Storage, privacy risk \\
\addlinespace
Feature & Shipped features with adoption below meaningful threshold &
Dev cost, complexity \\
\addlinespace
Attention & Interface elements that consume attention without
producing value & Cognitive capacity \\
\addlinespace
Decision & Flows requiring choices where a better default eliminates
the decision & Time, error rate \\
\addlinespace
AI & AI deployed where deterministic logic or simpler UX would
produce equivalent outcomes & Energy, latency \\
\addlinespace
Trust & Dark patterns or obscured behaviour that erodes trust over
time & Retention, reputation \\
\addlinespace
Cognitive & Complication exceeding the inherent demands of the task
\cite{norman2010} & Error rate, burnout \\
\bottomrule
\end{tabularx}
\end{table}

This taxonomy is not intended to be exhaustive. It is intended to be
memorable, actionable, and mutually distinguishable enough that product
teams can use it as an audit checklist. Future empirical work should
establish whether these eight categories are the right decomposition,
whether they are genuinely independent, and whether they reliably
predict measurable costs.

\subsection{Human-Centred Complexity}

Following Norman~\cite{norman2010}, GreenZ distinguishes
\textit{inherent complexity}, the irreducible structure of a domain,
which it is the designer's obligation to surface, not hide, from
\textit{interface complication}, which is complexity added by the
system beyond what the domain requires and which is always a design
failure. Healthcare triage genuinely involves many variables;
pretending otherwise is not sustainability, it is oversimplification.
The design task is to match interface complexity to domain complexity,
no more and no less.

\subsection{Responsible Intelligence}

GreenZ treats AI as simultaneously a liability and an asset. Intelligent
behaviour is powerful precisely because it makes autonomous judgments,
and that power carries proportional risk. The responsible intelligence
stance requires four things: every AI feature must be able to answer
what specific value it produces that a simpler approach cannot;
uncertainty must be surfaced, not hidden; human override paths must be
first-class design elements, not safety additions bolted on after
launch; and the resource cost of each AI interaction must be treated
as a design input, not an infrastructure detail left to engineering.

This stance synthesizes Amershi et al.~\cite{amershi2019},
Crawford~\cite{crawford2021}, Luccioni et al.~\cite{luccioni2024},
and emerging oversight-by-design literature.

\section{Layer 2: Operational Frameworks}

Five operational frameworks, each answering one question a product
team actually has to answer.

\subsection{Sustainable UX Audit Framework}
\textit{Where is waste and friction hiding?}

A structured method for identifying digital waste in existing systems.
Five phases: Alignment and Scoping (identify target workflows and agree
on what waste means for this system); System and Experience Mapping
(map workflows end-to-end, including people, systems, decisions, data
flows, and feedback loops); Waste Classification (classify identified
issues against the taxonomy in Table~\ref{tab:taxonomy}); Severity and
Cost Estimation (estimate the cost of each waste type using the Metrics
Framework); Roadmap Generation (prioritize by impact, effort, and risk).

Outputs: a visual audit report, redesign directions, a prioritized
roadmap, and a baseline measurement set.

\subsection{AI Sufficiency Decision Framework}
\label{sec:ai}
\textit{Should AI even be here?}

This is GreenZ's most distinctive contribution to the responsible AI
literature. Before applying Amershi et al.'s interaction guidelines or
Microsoft's HAX Toolkit, both of which assume AI is present, this
framework asks whether AI should be present at all.

The AI Sufficiency Decision Model proceeds through nine questions:

\begin{enumerate}[leftmargin=*, label=\arabic*.]
\item \textbf{What is the job?} State the specific outcome this feature
is supposed to produce.
\item \textbf{What happens if it goes wrong?} Classify the failure
mode: annoying, costly, dangerous, or consequential.
\item \textbf{Can deterministic logic solve this?} If the answer is
always the same given the same inputs, AI adds cost without adding value.
\item \textbf{Can better UX defaults solve this?} Smart defaults,
progressive disclosure, and clearer information architecture often
eliminate the need for AI assistance entirely.
\item \textbf{What specific value does AI add that alternatives cannot?}
Speed, consistency, personalization, anomaly detection, or
prediction, name it precisely. ``AI makes it smarter'' is not an
answer.
\item \textbf{Is the user better off knowing what the system is doing?}
Explainability is not optional when outcomes affect health, money,
safety, or rights.
\item \textbf{What is the compute and resource cost per interaction?}
Use Luccioni et al.'s~\cite{luccioni2024} energy benchmarks as
calibration. If this cannot be estimated, that is itself a design
problem.
\item \textbf{Who bears the cost if this goes wrong?} Map risk to the
stakeholders who actually experience it.
\item \textbf{What is the threshold at which this gets removed?} Define
it before shipping, not after.
\end{enumerate}

If questions 1--4 can be answered without AI, the default decision is:
do not deploy AI. The burden of proof is on the AI feature, not on its
absence.

\subsection{Complexity Reduction Framework}
\textit{How do we simplify without oversimplifying?}

Grounded in Norman~\cite{norman2010}, Endsley~\cite{endsley1995}, and
Klein~\cite{klein1999}, four principles guide the work. Match interface
structure to domain structure: do not flatten hierarchies that genuinely
exist, but do not add hierarchy that does not. Design for recognition
over comparison: Klein's Recognition-Primed Decision model shows experts
recognize patterns; give them patterns, not option menus. Support
projection, not just perception: Endsley's Level~3 situation awareness
is the level most enterprise UIs fail at; show consequences, not just
current state. Name the irreducible complexity: if something is
genuinely complicated, say so; hiding it produces false confidence and
downstream errors.

\subsection{Sustainable Metrics Framework}
\textit{How do we measure sustainable UX?}

GreenZ proposes a dual-axis measurement model: an efficiency axis
tracking user-facing outcomes, and a sufficiency axis tracking
system-facing resource use. Table~\ref{tab:metrics} lists the proposed
metrics.

\begin{table}[h]
\caption{GreenZ Metrics Framework}
\label{tab:metrics}
\small
\begin{tabularx}{\columnwidth}{p{1.8cm} p{0.9cm} X}
\toprule
Metric & Axis & What It Measures \\
\midrule
Task Success Rate & Eff. & Users completing key journeys without
critical errors \\
Time-to-Task & Eff. & Task completion time; proxy for friction \\
Attention Cost Index & Eff. & Interruptions and unsolicited
notifications per session \\
Calls per Key Flow & Suff. & API/AI calls per completed journey;
proxy for compute waste \\
Feature Adoption Rate & Suff. & \% of shipped features actively used \\
Data Collected vs. Used & Suff. & Fields captured versus fields that
influence decisions \\
AI Necessity Score & Suff. & \% of AI interactions where output was
accepted unmodified \\
Error/Escalation Rate & Both & Frequency of AI output correction or
human escalation \\
Carbon Shadow Estimate & Suff. & Approximate compute energy per task,
calibrated to \cite{luccioni2024} \\
\bottomrule
\end{tabularx}
\end{table}

A measurement caveat applies throughout. Carbon estimates derived from
API-call proxies and page-weight calculations carry significant
methodological uncertainty~\cite{debugbear2024}. These metrics are
designed to establish direction of travel, not precise absolute values.
Greenwashing risk is real; GreenZ names it rather than obscuring it.

\subsection{Responsible Workflow Framework}
\textit{How do we design for high-stakes systems?}

GreenZ treats every enterprise operational workflow as high-stakes by
default until evidence demonstrates otherwise. This disposition is
derived from aviation Crew Resource Management, patient-safety
literature~\cite{reason1990,gawande2009}, and emerging
oversight-by-design research.

Five structural requirements follow from this posture. Escalation paths
must be first-class UI elements, not buried settings. Kill switches
must precede launch: every AI-assisted decision pathway needs a visible,
accessible, fast human override. Failure states must be designed, not
discovered: error states and degraded modes receive the same design
attention as primary flows. Consequential actions require confirmation
at proportionate cost: not a modal for every click, but genuine
friction for irreversible or high-impact actions. System behaviour must
be explainable at the moment of decision, not buried in documentation.

\section{Layer 3: Tools and Canvases}

Layer~3 translates the operational frameworks into working artifacts:
a Digital Waste Audit Worksheet mapping identified issues to the eight
waste types with severity and cost fields; an AI Sufficiency Decision
Tree providing a visual flowchart of the nine-question model for use
in product-design reviews; a Complexity Heat Map Canvas for mapping
interface complication against domain complexity; a GreenZ Sprint
Template for applying the framework to a specific flow within two to
five days; and a Sustainable Metrics Dashboard Template covering both
axes.

Full versions of these tools are available at \url{uxperiment.design}.
Selected tools are released under Creative Commons Attribution 4.0 for
academic and non-commercial use.

\section{Worked Example: Applying GreenZ to an AI Triage Feature}

To illustrate how the framework operates in practice, we walk through
a representative enterprise scenario: a healthcare operations platform
considering whether to add an AI-powered patient triage suggestion
feature to its intake workflow.

\subsection{Context}

The platform is used by intake coordinators at a mid-sized outpatient
clinic. The existing workflow requires coordinators to read incoming
patient notes, assess urgency, and assign a triage category manually.
The product team is proposing an AI feature that reads the notes and
suggests a triage category before the coordinator reviews them. The
stated goal is to reduce intake time and improve consistency across
coordinators.

\subsection{Step 1: Sustainable UX Audit}

Before evaluating the AI feature, the audit framework maps the existing
workflow. The mapping surfaces three findings. First, coordinators
report that the intake form collects fifteen fields, but clinical
interviews reveal that only six reliably affect triage decisions,
nine fields are captured speculatively. This is Data Waste. Second,
the system requires coordinators to navigate four screens to complete
a single intake; two of those screens exist for historical reporting
reasons and are not used in the triage decision itself. This is
Decision Waste and Cognitive Waste. Third, the platform sends an
automated notification to the coordinator after each intake submission
confirming a task they just completed themselves. This is Attention
Waste.

None of these findings require AI to address. They represent
remediable waste in the existing system that should be resolved before
any new feature is added.

\subsection{Step 2: AI Sufficiency Decision Model}

With the audit complete, the team applies the nine-question model to
the proposed triage suggestion feature.

\textbf{Q1 --- What is the job?} Reduce the time coordinators spend
reading and categorizing intake notes, and reduce inter-coordinator
variability in triage decisions.

\textbf{Q2 --- What happens if it goes wrong?} A miscategorized
triage decision delays appropriate care. This failure mode is
consequential: it affects patient safety.

\textbf{Q3 --- Can deterministic logic solve this?} No. Triage
involves ambiguous, unstructured clinical language that varies by
patient. Rule-based systems have been tried in this domain and produce
unacceptable false-negative rates for high-acuity cases.

\textbf{Q4 --- Can better UX defaults solve this?} Partially.
Reducing the form from fifteen to six fields and removing the two
redundant screens will reduce cognitive load and likely improve
consistency on its own. The team should pilot these changes before
committing to AI. This is not a reason to reject AI permanently; it
is a reason to establish a baseline first.

\textbf{Q5 --- What specific value does AI add?} If the UX
improvements alone do not close the variability gap, AI adds
pattern-recognition across a larger volume of historical cases than
any individual coordinator has seen. That is a legitimate and
specific value proposition.

\textbf{Q6 --- Is the user better off knowing what the system is
doing?} Yes, unambiguously. The coordinator must be able to see which
fields drove the suggestion and disagree with the result. Unexplained
triage suggestions in a patient-safety context are not acceptable.
Explainability is a design requirement, not a nice-to-have.

\textbf{Q7 --- What is the compute and resource cost per interaction?}
The team estimates the feature will run inference on every intake
submission, approximately 800 per month. Using Luccioni et
al.'s~\cite{luccioni2024} energy benchmarks as a calibration
reference, the team logs this as a known and bounded cost, reviews
model options for task-specific rather than generative approaches, and
documents the decision.

\textbf{Q8 --- Who bears the cost if this goes wrong?} The patient,
in the first instance. The coordinator bears professional risk. The
clinic bears liability risk. This hierarchy of consequence determines
the design requirements: the coordinator must always be the decision
maker; the AI is advisory only.

\textbf{Q9 --- What is the removal threshold?} If the AI suggestion
is overridden by coordinators more than 30\% of the time after a
90-day pilot, the feature is reviewed for removal or replacement.
This threshold is documented before launch, not negotiated after
adoption has created inertia.

\subsection{Step 3: Complexity Reduction and Responsible Workflow}

The feature is scoped as an advisory suggestion, displayed inline with
the intake form, with the reasoning visible and a one-click override
requiring no additional confirmation. The override is logged silently,
coordinators are not asked to explain their disagreement, which
would add friction and create a chilling effect on overrides.

The kill switch is a toggle in the coordinator's own interface, not an
admin setting. Any coordinator can disable suggestions for their
session without contacting IT.

Failure states are designed before launch: if the model returns low
confidence, the suggestion is withheld entirely rather than displayed
with a confidence score that coordinators would anchor on regardless of
its value.

\subsection{Step 4: Metrics}

The team tracks four metrics from the GreenZ Metrics Framework: Task
Success Rate (intake completion without escalation), Time-to-Task
(mean intake time before and after), AI Necessity Score (percentage of
suggestions accepted unmodified, as a proxy for appropriate calibration,
too high suggests over-reliance; too low suggests the feature
provides no value), and Error/Escalation Rate (cases subsequently
re-triaged by a senior clinician).

\subsection{What This Example Demonstrates}

The GreenZ analysis of this scenario produces four outputs that
existing frameworks do not. It identifies and addresses waste in the
existing system before adding new capability. It establishes that AI
is justified in this case but only after simpler interventions are
trialled. It produces specific, non-negotiable design requirements
from the sufficiency questions, explainability, coordinator
override, documented removal threshold, rather than treating these
as optional quality attributes. And it names the metrics and the
removal condition before a single line of model code is written.

This sequence - audit first, AI sufficiency second, complexity
reduction third, metrics fourth - is the intended operating order
of GreenZ in practice.

\section{Positioning, Limitations, and Future Work}

\subsection{Positioning Relative to Existing Frameworks}

GreenZ is not a replacement for existing frameworks. It is designed to
work alongside them. It extends Nielsen's heuristics by adding
waste-detection and sufficiency criteria to usability evaluation. It
precedes Amershi et al.'s 18 guidelines by adding a prior question:
should AI be here? It operationalizes Santarius et al.'s digital
sufficiency for product teams who are not academics. It anticipates the
W3C WSG by providing an enterprise-grade reading of sustainability
guidelines before they reach W3C Recommendation status, targeted for
Earth Day 2026. Table~\ref{tab:positioning} summarizes GreenZ's
position against the most relevant existing frameworks.

\begin{table}[h]
\caption{GreenZ Positioning Against Existing Frameworks}
\label{tab:positioning}
\small
\begin{tabularx}{\columnwidth}{X c c c c c}
\toprule
Dimension & \rotatebox{60}{Nielsen} & \rotatebox{60}{HEART} &
\rotatebox{60}{Amershi} & \rotatebox{60}{WSG} & \rotatebox{60}{\textbf{GreenZ}} \\
\midrule
Sustainability criteria    & -- & -- & -- & \checkmark & \checkmark \\
AI necessity question      & -- & -- & -- & -- & \checkmark \\
Cognitive sufficiency      & $\sim$ & -- & $\sim$ & -- & \checkmark \\
Enterprise/high-stakes     & -- & $\sim$ & -- & -- & \checkmark \\
Digital waste taxonomy     & -- & -- & -- & -- & \checkmark \\
Measurement model          & -- & \checkmark & $\sim$ & \checkmark & \checkmark \\
Open tools layer           & -- & -- & \checkmark & \checkmark & \checkmark \\
Empirically validated      & \checkmark & \checkmark & \checkmark & $\sim$ & -- \\
\bottomrule
\multicolumn{6}{l}{\footnotesize \checkmark~=~yes; $\sim$~=~partial; --~=~no}
\end{tabularx}
\end{table}

\subsection{Limitations}

\textbf{This framework has not yet been empirically validated.} The
Digital Waste Taxonomy, the AI Sufficiency Decision Model, and the
Metrics Framework are derived from synthesis of existing literature
and practitioner observation. The intended next step, a structured
expert review study recruiting five to eight senior practitioners
across enterprise UX, product management, and AI product
development, is underway at time of submission. Participants are
evaluating each framework component for face validity, internal
consistency, and perceived usefulness. Results will inform v2.
Reviewers should treat this paper as a framework proposal inviting
critique, not a validated instrument.

\textbf{Sustainability metrics are directional, not precise.}
Carbon shadow estimates derived from API-call proxies carry significant
methodological uncertainty~\cite{debugbear2024}. The Metrics Framework
is designed as a directional signal system. Teams using GreenZ metrics
should disclose their measurement methodology and avoid presenting
estimates as precise measurements.

\textbf{Design Degrowth is a design-process discipline, not a political
prescription.} The degrowth theoretical tradition~\cite{latouche2009,
hickel2020} operates at the level of political economy and carries
contested ideological associations. GreenZ's application of sufficiency
is a constraint within the design process, not a prescription for
organizational growth strategy. An enterprise pursuing aggressive
revenue growth can apply this principle consistently within its design
decisions. These are not the same question.

\textbf{Cross-cultural validation is a stated gap, not an oversight.}
The Cultural Ecosystem Equity principle exists precisely because the
framework's tools have not been validated across cultural contexts.
The audit instruments reflect a predominantly Western enterprise
software context. Cross-cultural validation is named as a Phase~2
research priority.

\subsection{Future Work}

Four empirical directions follow from this paper. A practitioner
validation study is underway, assessing face validity and perceived
usefulness across enterprise UX, product management, and AI product
development contexts. A Digital Waste Taxonomy refinement study using
card-sorting and expert review will assess whether the eight categories
are the right decomposition, mutually exclusive, and collectively
exhaustive. Longitudinal case studies will measure task efficiency,
feature adoption, and compute-cost outcomes before and after GreenZ
adoption. And a CHI 2027 submission will position GreenZ within the
WCAG-to-WSG standardization trajectory, arguing for a formal enterprise
sustainability standard in the WSG tradition.

GreenZ is versioned. This is v1. Updates will be published at
\url{uxperiment.design} and via arXiv revision as empirical evidence
accumulates.

\section{Conclusion}

Digital systems will not become more sustainable by accident. The
structural incentives of the software industry - feature velocity,
engagement maximization, AI capability signalling - produce waste as a
default outcome. Reversing this requires frameworks that name the
problem with precision, give practitioners tools they can use, and
build a body of evidence capable of influencing standards bodies and
enterprise procurement decisions.

GreenZ is an attempt to begin that accumulation. It synthesizes two
decades of sustainable HCI research, practitioner frameworks for AI
responsibility, cognitive science of enterprise decision-making, and
degrowth theory into a three-layer system that a product team can use
on a Monday morning and a researcher can build on in a CHI submission.

The central claim is simple: sufficiency is a design criterion, not an
afterthought. What is not needed should not be built. What is built
should not waste. What wastes should be named, measured, and removed.

GreenZ is a working document, not a finished one. The field does not
need more frameworks that claim completeness. It needs frameworks that
invite critique, absorb evidence, and improve. This one is designed
to do all three.

\bibliographystyle{ACM-Reference-Format}

\appendix

\section{GreenZ Principles and Operational Frameworks}
\label{app:mapping}

\begin{table}[h]
\caption{Principle-to-Framework Mapping}
\small
\begin{tabularx}{\columnwidth}{X p{1.8cm} p{1.8cm}}
\toprule
GreenZ Principle & Primary Framework & Secondary Framework \\
\midrule
Regenerative Interfaces & Responsible Workflow & SUX Audit \\
Waste-Conscious UX & SUX Audit & Metrics \\
Adaptive Systems Thinking & Complexity Reduction & --- \\
Carbon Shadow Audits & Metrics & AI Sufficiency \\
Psychologically Regenerative Design & Complexity Reduction & Responsible Workflow \\
Cultural Ecosystem Equity & Responsible Workflow & SUX Audit \\
Design Degrowth & AI Sufficiency & Complexity Reduction \\
Time-to-Value Conservation & Metrics & SUX Audit \\
Regenerative Design Systems & SUX Audit & --- \\
Speculative Sustainability R\&D & Future work & All \\
\bottomrule
\end{tabularx}
\end{table}

\vspace{6pt}
\noindent\textit{\textcopyright~2026 UXperiment Inc. This preprint is
shared under Creative Commons Attribution 4.0 International (CC~BY~4.0).
Cite as: Solanki, T. (2026). GreenZ: A Sustainable UX Framework for
Complex Digital Systems. arXiv preprint.}

\end{document}